\documentclass[runningheads]{svmult}

\usepackage{makeidx}   

\usepackage{graphicx}  

\usepackage{subeqnar}  

\usepackage{multicol}  

\usepackage{cropmark}

\usepackage{physprbb}

\begin{document}

\title*{Cosmoparticle Physics -\protect\newline the Challenge for the
Millenium}

\toctitle{Cosmoparticle Physics -

\protect\newline the Challenge for the Millenium}


%

%

\titlerunning{Cosmoparticle Physics}



%

\author{Maxim Yu. Khlopov \inst{1,2}}

\authorrunning{Maxim Khlopov}

\institute{ Center for Cosmoparticle Physics "Cosmion", 125047,
Moscow, Russia \and Physics Department, University "LaSapienza",
Ple A.Moro,2, 00185, Rome, Italy}

\maketitle              

\begin{abstract}

Cosmoparticle physics is the natural result of development of
mutual relationship between cosmology and particle physics. Its
prospects offer the way to study the theory of everything and the
true history of the Universe, based on it, in the proper
combination of their indirect physical, astrophysical and
cosmological signatures. We may be near the first positive results
in this direction. The basic ideas of cosmoparticle physics are
briefly reviewed.

\end{abstract}

Cosmoparticle physics originates from the well established
relationship between microscopic and macroscopic descriptions in
theoretical physics. Remind the links between statistical physics
and thermodynamics, or between electrodynamics and theory of
electron. To the end of the XX Century the new level of this
relationship was realized. It followed both from the cosmological
necessity to go beyond the world of known elementary particles in
the physical grounds for inflationary cosmology with
baryosynthesis and dark matter as well as from the necessity for
particle theory to use cosmological tests as the important and in
many cases unique way to probe its predictions.

The convergence of the frontiers of our knowledge in micro- and
macro worlds leads to the wrong circle of problems, illustrated by
the mystical Uhroboros (self-eating-snake). The Uhroboros puzzle
may be formulated as follows: {\it The theory of the Universe is
based on the predictions of particle theory, that need cosmology
for their test}. Cosmoparticle physics \cite{ADS}, \cite{MKH},
\cite{book} offers the way our of this wrong circle. It studies
the fundamental basis and mutual relationship between micro-and
macro-worlds in the proper combination of physical, astrophysical
and cosmological signatures.

Let's specify in more details the set of links between fundamental
particle properties and their cosmological effects.

The role of particle content in the Einstein equations is reduced
to its contribution into energy-momentum tensor. So, the set of
relativistic species, dominating in the Universe, realizes the
relativistic equation of state $p= \varepsilon/3$ and the
relativistic stage of expansion. The difference between
relativistic bosons and fermions or various bosonic (or fermionic)
species is accounted by the statistic weight of respective degree
of freedom. The very treatment of different species of particles
as equivalent degrees of freedom physically assumes strict
symmetry between them.

Such strict symmetry is not realized in Nature. There is no exact
symmetry between bosons and fermions (e.g. supersymmetry). There
is no exact symmetry between various quarks and leptons. The
symmetry breaking implies the difference in particle masses. The
particle mass pattern reflects the hierarchy of symmetry breaking.

Noether's theorem relates the exact symmetry to conservation of
respective charge. The lightest particle, bearing the strictly
conserved charge, is absolutely stable. So, electron is absolutely
stable, what reflects the conservation of electric charge. In the
same manner the stability of proton is conditioned by the
conservation of baryon charge. The stability of ordinary matter is
thus protected by the conservation of electric and baryon charges,
and its properties reflect the fundamental physical scales of
electroweak and strong interactions. Indeed, the mass of electron
is related to the scale of the electroweak symmetry breaking,
whereas the mass of proton reflects the scale of QCD confinement.

Extensions of the standard model imply new symmetries and new
particle states. The respective symmetry breaking induces new
fundamental physical scales in particle theory. If the symmetry is
strict, its existence implies new conserved charge. The lightest
particle, bearing this charge, is stable. The set of new
fundamental particles, corresponding to the new strict symmetry,
is then reflected in the existence of new stable particles, which
should be present in the Universe and taken into account in the
total energy-momentum tensor.

Most of the known particles are unstable. For a particle with the
mass $m$ the particle physics time scale is $t \sim 1/m$, so in
particle world we refer to particles with lifetime $\tau \gg 1/m$
as to metastable. To be of cosmological significance metastable
particle should survive after the temperature of the Universe $T$
fell down below $T \sim m$, what means that the particle lifetime
should exceed $t \sim (m_{Pl}/m) \cdot (1/m)$. Such a long
lifetime should find reason in the existence of an (approximate)
symmetry. From this viewpoint, cosmology is sensitive to the most
fundamental properties of microworld, to the conservation laws
reflecting strict or nearly strict symmetries of particle theory.

However, the mechanism of particle symmetry breaking can also have
the cosmological impact. Heating of condensed matter leads to
restoration of its symmetry. When the heated matter cools down,
phase transition to the phase of broken symmetry takes place. In
the course of the phase transitions, corresponding to given type
of symmetry breaking, topological defects can form. One can
directly observe formation of such defects in liquid crystals or
in superfluids. In the same manner the mechanism of spontaneous
breaking of particle symmetry  implies restoration of the
underlying symmetry. When temperature decreases in the course of
cosmological expansion, transitions to the phase of broken
symmetry  can lead, depending on the symmetry breaking pattern, to
formation of topological defects in very early Universe. The
defects can represent the new form of stable particles (as it is
in the case of magnetic monopoles), or the form of extended
structures, such as cosmic strings or cosmic walls.

In the old Big bang scenario the cosmological expansion and its
initial conditions was given {\it a priori}. In the modern
cosmology the expansion of the Universe and its initial conditions
is related to the process of inflation. The global properties of
the Universe as well as the origin of its large scale structure
are the result of this process. The matter content of the modern
Universe is also originated from the physical processes: the
baryon density is the result of baryosynthesis and the nonbaryonic
dark matter represents the relic species of physics of the hidden
sector of particle theory. Physics, underlying inflation,
baryosynthesis and dark matter, is referred to the extensions of
the standard model, and the variety of such extensions makes the
whole picture in general ambiguous. However, in the framework of
each particular physical realization of inflationary model with
baryosynthesis and dark matter the corresponding model dependent
cosmological scenario can be specified in all the details. In such
scenario the main stages of cosmological evolution, the structure
and the physical content of the Universe reflect the structure of
the underlying physical model. The latter should include with
necessity the standard model, describing the properties of
baryonic matter, and its extensions, responsible for inflation,
baryosynthesis and dark matter. In no case the cosmological impact
of such extensions is reduced to reproduction of these three
phenomena only. The nontrivial path of cosmological evolution,
specific for each particular realization of inflational model with
baryosynthesis and nonbaryonic dark matter, always contains some
additional model dependent cosmologically viable predictions,
which can be confronted with astrophysical data. The part of
cosmoparticle physics, called cosmoarcheology, offers the set of
methods and tools probing such predictions.

Cosmoarcheology considers the results of observational cosmology
as the sample of the experimental data on the possible existence
and features of hypothetical phenomena predicted by particle
theory. To undertake the {\it Gedanken Experiment} with these
phenomena some theoretical framework to treat their origin and
evolution in the Universe should be assumed. As it was pointed out
in \cite{Cosmoarcheology} the choice of such framework is a
nontrivial problem in the modern cosmology.

Indeed, in the old Big bang scenario any new phenomenon, predicted
by particle theory was considered in the course of the thermal
history of the Universe, starting from Planck times. The problem
is that the bedrock of the modern cosmology, namely, inflation,
baryosynthesis and dark matter, is also based on experimentally
unproven part of particle theory, so that the test for possible
effects of new physics is accomplished by the necessity to choose
the physical basis for such test. There are two possible solutions
for this problem: a) a crude model independent comparison of the
predicted effect with the observational data and b) the model
dependent treatment of considered effect, provided that the model,
predicting it, contains physical mechanism of inflation,
baryosynthesis and dark matter.

The basis for the approach (a) is that whatever happened in the
early Universe its results should not contradict the observed
properties of the modern Universe. The set of observational data
and, especially, the light element abundance and thermal spectrum
of microwave background radiation put severe constraint on the
deviation from thermal evolution after 1 s of expansion, what
strengthens the model independent conjectures of approach (a).

One can specify the new phenomena by their net contribution into
the cosmological density and by forms of their possible influence
on parameters of matter and radiation. In the first aspect we can
consider strong and weak phenomena. Strong phenomena can put
dominant contribution into the density of the Universe, thus
defining the dynamics of expansion in that period, whereas the
contribution of weak phenomena into the total density is always
subdominant. The phenomena are time dependent, being characterized
by their time-scale, so that permanent (stable) and temporary
(unstable) phenomena can take place. They can have homogeneous and
inhomogeneous distribution in space. The amplitude of density
fluctuations $\delta \equiv \delta \varrho/\varrho$ measures the
level of inhomogeneity relative to the total density, $\varrho$.
The partial amplitude $\delta_{i} \equiv \delta
\varrho_{i}/\varrho_{i}$ measures the level of fluctuations within
a particular component with density $\varrho_{i}$, contributing
into the total density $\varrho = \sum_{i} \varrho_{i}$. The case
$\delta_{i} \ge 1$ within the considered $i$-th component
corresponds to its strong inhomogeneity. Strong inhomogeneity is
compatible with the smallness of total density fluctuations, if
the contribution of inhomogeneous component into the total density
is small: $\varrho_{i} \ll \varrho$, so that $\delta \ll 1$.

The phenomena can influence the properties of matter and radiation
either indirectly, say, changing of the cosmological equation of
state, or via direct interaction with matter and radiation. In the
first case only strong phenomena are relevant, in the second case
even weak phenomena are accessible to observational data. The
detailed analysis of sensitivity of cosmological data to various
phenomena of new physics are presented in \cite{book}.

The basis for the approach (b) is provided by a particle model, in
which inflation, baryosynthesis and nonbaryonic dark matter is
reproduced. Any realization of such physically complete basis for
models of the modern cosmology contains with necessity additional
model dependent predictions, accessible to cosmoarcheological
means. Here the scenario should contain all the details, specific
to the considered model, and the confrontation with the
observational data should be undertaken in its framework. In this
approach complete cosmoparticle physics models may be realized,
where all the parameters of particle model can be fixed from the
set of astrophysical, cosmological and physical constraints. Even
the details, related to cosmologically irrelevant predictions,
such as the parameters of unstable particles, can find the
cosmologically important meaning in these models. So, in the model
of horizontal unification \cite{Berezhiani1}, \cite{Berezhiani2},
\cite{Sakharov1}, the top quark or B-meson physics fixes the
parameters, describing the dark matter, forming the large scale
structure of the Universe.

To study the imprints of new physics in astrophysical data
cosmoarcheology implies the forms and means in which new physics
leaves such imprints. So, the important tool of cosmoarcheology in
linking the cosmological predictions of particle theory to
observational data is the {\it Cosmophenomenology} of new physics.
It studies the possible hypothetical forms of new physics, which
may appear as cosmological consequences of particle theory, and
their properties, which can result in observable effects.

The simplest primordial form of new physics is the gas of new
stable massive particles, originated from early Universe. For
particles with the mass $m$, at high temperature $T>m$ the
equilibrium condition, $n \cdot \sigma v \cdot t > 1$ is valid, if
their annihilation cross section $\sigma > 1/(m m_{Pl})$ is
sufficiently large to establish the equilibrium. At $T<m$ such
particles go out of equilibrium and their relative concentration
freezes out. More weakly interacting species decouple from plasma
and radiation at $T>m$, when $n \cdot \sigma v \cdot t \sim 1$,
i.e. at $T_{dec} \sim (\sigma m_{Pl})^{-1}$. The maximal
temperature, which is reached in inflationary Universe, is the
reheating temperature, $T_{r}$, after inflation. So, the very
weakly interacting particles with the annihilation cross section
$\sigma < 1/(T_{r} m_{Pl})$, as well as very heavy particles with
the mass $m \gg T_{r}$ can not be in thermal equilibrium, and the
detailed mechanism of their production should be considered to
calculate their primordial abundance.

Decaying particles with the lifetime $\tau$, exceeding the age of
the Universe, $t_{U}$, $\tau > t_{U}$, can be treated as stable.
By definition, primordial stable particles survive to the present
time and should be present in the modern Universe. The net effect
of their existence is given by their contribution into the total
cosmological density. They can dominate in the total density being
the dominant form of cosmological dark matter, or they can
represent its subdominant fraction. In the latter case more
detailed analysis of their distribution in space, of their
condensation in galaxies, of their capture by stars, Sun and
Earth, as well as of the effects of their interaction with matter
and of their annihilation provides more sensitive probes for their
existence. In particular, hypothetical stable neutrinos of the 4th
generation with the mass about 50 GeV are predicted to form the
subdominant form of the modern dark matter, contributing less than
0,1 \% to the total density. However, direct experimental search
for cosmic fluxes of weakly interacting massive particles (WIMPs)
may be sensitive to the existence of such component \cite{Cline},
\cite{Bernabei}, and may be even favors it \cite{Bernabei}. It was
shown in \cite{Fargion99}, \cite{Grossi}, \cite{Belotsky} that
annihilation of 4th neutrinos and their antineutrinos in the
Galaxy can explain the galactic gamma-background, measured by
EGRET in the range above 1 GeV, and that it can give some clue to
explanation of cosmic positron anomaly, claimed to be found by
HEAT. 4th neutrino annihilation inside the Earth should lead to
the flux of underground monochromatic neutrinos of known types,
which can be traced in the analysis of the already existing and
future data of underground neutrino detectors \cite{Belotsky}.

New particles with electric charge and/or strong interaction can
form anomalous atoms and contain in the ordinary matter as
anomalous isotopes. For example, if the lightest quark of 4th
generation is stable, it can form stable charged hadrons, serving
as nuclei of anomalous atoms of e.g. crazy helium \cite{BKS}.

Primordial unstable particles with the lifetime, less than the age
of the Universe, $\tau < t_{U}$, can not survive to the present
time. But, if their lifetime is sufficiently large to satisfy the
condition $\tau \gg (m_{Pl}/m) \cdot (1/m)$, their existence in
early Universe can lead to direct or indirect traces. Cosmological
flux of decay products contributing into the cosmic and gamma ray
backgrounds represents the direct trace of unstable particles. If
the decay products do not survive to the present time their
interaction with matter and radiation can cause indirect trace in
the light element abundance or in the fluctuations of thermal
radiation. If the particle lifetime is much less than $1$s the
multi-step indirect traces are possible, provided that particles
dominate in the Universe before their decay. On the dust-like
stage of their dominance black hole formation takes place, and the
spectrum of such primordial black holes traces the particle
properties (mass, frozen concentration, lifetime) \cite{Polnarev}.
The particle decay in the end of dust like stage influences the
baryon asymmetry of the Universe. In any way cosmophenomenoLOGICAL
chains link the predicted properties of even unstable new
particles to the effects accessible in astronomical observations.
Such effects may be important in the analysis of the observational
data.

So, the only direct evidence for the accelerated expansion of the
modern Universe comes from the distant SN I data. The data on the
cosmic microwave background (CMB) radiation and large scale
structure (LSS) evolution (see e.g. \cite{WMAP}) prove in fact the
existence of homogeneously distributed dark energy and the slowing
down of LSS evolution at $z \leq 3$. Homogeneous negative pressure
medium ($\Lambda$-term or quintessence) leads to {\it relative}
slowing down of LSS evolution due to acceleration of cosmological
expansion. However, both homogeneous component of dark matter and
slowing down of LSS evolution naturally follow from the models of
Unstable Dark Matter (UDM) (see \cite{book} for review), in which
the structure is formed by unstable weakly interacting particles.
The weakly interacting decay products are distributed
homogeneously. The loss of the most part of dark matter after
decay slows down the LSS evolution. The dominantly invisible decay
products can contain small ionizing component \cite{Berezhiani2}.
Thus, UDM effects will deserve attention, even if the accelerated
expansion is proved.

The parameters of new stable and metastable particles are also
determined by the pattern of particle symmetry breaking. This
pattern is reflected in the succession of phase transitions in the
early Universe. The phase transitions of the first order proceed
through the bubble nucleation, which can result in black hole
formation. The phase transitions of the second order can lead to
formation of topological defects, such as walls, string or
monopoles. The observational data put severe constraints on
magnetic monopole and cosmic wall production, as well as on the
parameters of cosmic strings. The succession of phase transitions
can change the structure of cosmological defects. The more
complicated forms, such as walls-surrounded-by-strings can appear.
Such structures can be unstable, but their existence can lead the
trace in the nonhomogeneous distribution of dark matter and in
large scale correlations in the nonhomogeneous dark matter
structures, such as {\it archioles} \cite{Sakharov2}. The large
scale correlations in topological defects and their imprints in
primordial inhomogeneities is the indirect effect of inflation, if
phase transitions take place after reheating of the Universe.
Inflation provides in this case the equal conditions of phase
transition, taking place in causally disconnected regions.

If the phase transitions take place on inflational stage new forms
of primordial large scale correlations appear. The example of
global U(1) symmetry, broken spontaneously in the period of
inflation and successively broken explicitly after reheating, was
recently considered in \cite{KRS}. In this model, spontaneous U(1)
symmetry breaking at inflational stage is induced by the vacuum
expectation value $\langle \psi \rangle = f$ of a complex scalar
field $\Psi = \psi \exp{(\I \theta)}$, having also explicit
symmetry breaking term in its potential $V_{eb} = \Lambda^{4} (1 -
\cos{\theta})$. The latter is negligible in the period of
inflation, if $f \gg \Lambda$, so that there appears a valley
relative to values of phase in the field potential in this period.
Fluctuations of the phase $\theta$ along this valley, being of the
order of $\Delta \theta \sim H/(2\pi f)$ (here $H$ is the Hubble
parameter at inflational stage) change in the course of inflation
its initial value within the regions of smaller size. Owing to
such fluctuations, for the fixed value of $\theta_{60}$ in the
period of inflation with {\it e-folding} $N=60$ corresponding to
the part of the Universe within the modern cosmological horizon,
strong deviations from this value appear at smaller scales,
corresponding to later periods of inflation with $N < 60$. If
$\theta_{60} < \pi$, the fluctuations can move the value of
$\theta_{N}$ to $\theta_{N} > \pi$ in some regions of the
Universe. After reheating, when the Universe cools down to
temperature $T = \Lambda$ the phase transition to the true vacuum
states, corresponding to the minima of $V_{eb}$ takes place. For
$\theta_{N} < \pi$ the minimum of $V_{eb}$ is reached at
$\theta_{vac} = 0$, whereas in the regions with $\theta_{N} > \pi$
the true vacuum state corresponds to $\theta_{vac} = 2\pi$. For
$\theta_{60} < \pi$ in the bulk of the volume within the modern
cosmological horizon $\theta_{vac} = 0$. However, within this
volume there appear regions with $\theta_{vac} = 2\pi$. These
regions are surrounded by massive domain walls, formed at the
border between the two vacua. Since regions with $\theta_{vac} =
2\pi$ are confined, the domain walls are closed. After their size
equals the horizon, closed walls can collapse into black holes.
The minimal mass of such black hole is determined by the condition
that it's Schwarzschild radius, $r_{g} = 2 G M/c^{2}$ exceeds the
width of the wall, $l \sim f/\Lambda^{2}$, and it is given by
$M_{min} \sim f (m_{Pl}/\Lambda)^{2}$. The maximal mass is
determined by the mass of the wall, corresponding to the earliest
region $\theta_{N} > \pi$, appeared at inflational stage.  This
mechanism can lead to formation of primordial black holes of a
whatever large mass (up to the mass of AGNs \cite{AGN}). Such
black holes appear in the form of primordial black hole clusters,
exhibiting fractal distribution in space \cite{KRS}. It can shed
new light on the problem of galaxy formation.

Primordial strong inhomogeneities can also appear in the baryon
charge distribution. The appearance of antibaryon domains in the
baryon asymmetrical Universe, reflecting the inhomogeneity of
baryosynthesis, is the profound signature of such strong
inhomogeneity \cite{CKSZ}. On the example of the model of
spontaneous baryosynthesis (see \cite{Dolgov} for review) the
possibility for existence of antimatter domains, surviving to the
present time in inflationary Universe with inhomogeneous
baryosynthesis was revealed in \cite{KRS2}. Evolution of
sufficiently dense antimatter domains can lead to formation of
antimatter globular clusters \cite{GC}. The existence of such
cluster in the halo of our Galaxy should lead to the pollution of
the galactic halo by antiprotons. Their annihilation can reproduce
\cite{Golubkov} the observed galactic gamma background in the
range tens-hundreds MeV. The prediction of antihelium component of
cosmic rays \cite{ANTIHE}, accessible to future searches for
cosmic ray antinuclei in PAMELA and AMS II experiments, as well as
of antimatter meteorites \cite{ANTIME} provides the direct
experimental test for this hypothesis.

So the primordial strong inhomogeneities in the distribution of
total, dark matter and baryon density in the Universe is the new
important phenomenon of cosmological models, based on particle
models with hierarchy of symmetry breaking.

The new physics follows from the necessity to extend the Standard
model. The white spots in the representations of symmetry groups,
considered in the extensions of the Standard model, correspond to
new unknown particles. The extension of the symmetry of gauge
group puts into consideration new gauge fields, mediating new
interactions. Global symmetry breaking results in the existence of
Goldstone boson fields.

For a long time the necessity to extend the Standard model had
purely theoretical reasons. Aesthetically, because full
unification is not achieved in the Standard model; practically,
because it contains some internal inconsistencies. It does not
seem complete for cosmology. One has to go beyond the Standard
model to explain inflation, baryosynthesis and nonbaryonic dark
matter. Recently there has appeared a set of experimental
evidences for the existence of neutrino oscillations (see for
recent review e.g. \cite{Ishihara}, \cite{Ewan}, \cite{Jung}), of
cosmic WIMPs \cite{Bernabei}, and of double neutrinoless beta
decay \cite{Klapdor}. Whatever is the accepted status of these
evidences, they indicate that the experimental searches may have
already crossed the border of new physics.

In particle physics direct experimental probes for the predictions
of particle theory are most attractive. The predictions of new
charged particles, such as supersymmetric particles or quarks and
leptons of new generation, are accessible to experimental search
at accelerators of new generation, if their masses are in
100GeV-1TeV range. However, the predictions related to higher
energy scale need non-accelerator or indirect means for their
test.

The search for rare processes, such as proton decay, neutrino
oscillations, neutrinoless beta decay, precise measurements of
parameters of known particles, experimental searches for dark
matter represent the widely known forms of such means.
Cosmoparticle physics offers the nontrivial extensions of indirect
and non-accelerator searches for new physics and its possible
properties. In experimental cosmoarcheology the data is to be
obtained, necessary to link the cosmophenomenology of new physics
with astrophysical observations (See \cite{Cosmoarcheology}). In
experimental cosmoparticle physics the parameters, fixed from the
consitency of cosmological models and observations, define the
level, at which the new types of particle processes should be
searched for (see \cite{expcpp}).

The theories of everything should provide the complete physical
basis for cosmology. The problem is that the string theory
\cite{Green} is now in the form of "theoretical theory", for which
the experimental probes are widely doubted to exist. The
development of cosmoparticle physics can remove these doubts. In
its framework there are two directions to approach the test of
theories of everything.

One of them is related with the search for the experimentally
accessible effects of heterotic string phenomenology. The
mechanism of compactification and symmetry breaking leads to the
prediction of homotopically stable objects \cite{Kogan1} and
shadow matter \cite{Kogan2}, accessible to cosmoarcheological
means of cosmoparticle physics. The condition to reproduce the
Standard model naturally leads in the heterotic string
phenomenology to the prediction of fourth generation of quarks and
leptons \cite{Shibaev} with a stable massive 4th neutrino
\cite{Fargion99}, what can be the subject of complete experimental
test in the near future. The comparison between the rank of the
unifying group $E_{6}$ ($r=6$) and the rank of the Standard model
($r=4$) implies the existence of new conserved charges and new
(possibly strict) gauge symmetries. New strict gauge U(1) symmetry
(similar to U(1) symmetry of electrodynamics) is possible, if it
is ascribed to the fermions of 4th generation. This hypothesis
explains the difference between the three known types of neutrinos
and neutrino of 4th generation. The latter possesses new gauge
charge and, being Dirac particle, can not have small Majorana mass
due to sea saw mechanism. If the 4th neutrino is the lightest
particle of the 4th quark-lepton family, strict conservation of
the new charge makes massive 4th neutrino to be absolutely stable.
Following this hypothesis \cite{Shibaev} quarks and leptons of 4th
generation are the source of new long range interaction
($y$-electromagnetism), similar to the electromagnetic interaction
of ordinary charged particles. New strictly conserved local U(1)
gauge symmetries can also arise in the development of D-brane
phenomenology \cite{Alday1}, \cite{Alday2}. If proved, the
practical importance of this property could be hardly
overestimated.

It is interesting, that heterotic string phenomenology  embeds
even in its simplest realisation both supersymmetric particles and
the 4th family of quarks and leptons, in particular, the two types
of WIMP candidates: neutralinos and massive stable 4th neutrinos.
So in the framework of this phenomenology the multicomponent
analysis of WIMP effects is favorable.

In the above approach some particular phenomenological features of
simplest variants of string theory are studied. The other
direction is to elaborate the extensive phenomenology of theories
of everything by adding to the symmetry of the Standard model the
(broken) symmetries, which have serious reasons to exist. The
existence of (broken) symmetry between quark-lepton families, the
necessity in the solution of strong CP-violation problem with the
use of broken Peccei-Quinn symmetry, as well as the practical
necessity in supersymmetry to eliminate the quadratic divergence
of Higgs boson mass in electroweak theory is the example of
appealing additions to the symmetry of the Standard model. The
horizontal unification and its cosmology represent the first step
on this way, illustrating the approach of cosmoparticle physics to
the elaboration of the proper phenomenology for theories of
everything \cite{Sakharov1}.

We can conclude that from the very beginning to the modern stage,
the evolution of Universe is governed by the forms of matter,
different from those we are built of and observe around us. From
the very beginning to the present time, the evolution of the
Universe was governed by physical laws, which we still don't know.
Observational cosmology offers strong evidences favoring the
existence of processes, determined by new physics, and the
experimental physics approaches to their investigation.

Cosmoparticle physics \cite{ADS} \cite{MKH}, studying the
physical, astrophysical and cosmological impact of new laws of
Nature, explores the new forms of matter and their physical
properties, what opens the way to use the corresponding new
sources of energy and new means of energy transfer. It offers the
great challenge for the new Millennium.

The work was performed in the framework of the State Contract
40.022.1.1.1106 and was partially supported by the RFBR grant
02-02-17490 and grant UR.02.01.026.

\end{document}